\documentclass[11pt,twoside]{article}  % Leave intact
\usepackage{apn3conf}

\begin{document}   % Leave intact

%-----------------------------------------------------------------------
%		            Paper Title 
%-----------------------------------------------------------------------
% Enter the title of the paper.
%
% EXAMPLE: \title{A Breakthrough in Astronomical Software Development}
%
% If your title is so long as to fill the page header when you print it,
% then please supply a short form as a \titlemark.
%
% EXAMPLE:
%  \title{Rapid Development for Distributed Computing, with Implications
%         for the Virtual Observatory}
%  \titlemark{Rapid Development for Distributed Computing}
%

\title{Binarity and Symbiotics in Asymmetrical Planetary Nebulae}
\titlemark{Binarity and Symbiotics}

%-----------------------------------------------------------------------
%		          Authors of Paper
%-----------------------------------------------------------------------
% Enter the authors followed by their affiliations.  The \author and
% \affil commands may appear multiple times as necessary.  List each
% author by giving the first name or initials first followed by the
% last name.  Authors with the same affiliations should grouped
% together. 
%
% Try to limit the front matter to no more than three \author
% commands.  Group authors with the same affiliations.  Too many
% \author commands fills the first page of the paper with little
% actual text.

\author{Hugo E. Schwarz, Hektor Monteiro}
\affil{Cerro Tololo Inter-American Observatory, NOAO-AURA, Casilla 603, 
La Serena, Chile.}

% Notice that some of these authors have alternate affiliations, which
% are identified by the \altaffilmark after each name.  The actual alternate
% affiliation information is typeset in footnotes at the bottom of the
% first page, and the text itself is specified in \altaffiltext commands.
% There is a separate \altaffiltext for each alternate affiliation
% indicated above.

%-----------------------------------------------------------------------
%			 Contact Information
%-----------------------------------------------------------------------
% This information will not appear in the paper but will be used by
% the editors in case you need to be contacted concerning your
% submission.  Enter your name as the contact along with your email
% address.

\contact{Hugo E. Schwarz}
\email{hschwarz@ctio.noao.edu}

%-----------------------------------------------------------------------
%		      Author Index Specification
%-----------------------------------------------------------------------
% Specify how each author name should appear in the author index.  The 
% \paindex{ } should be used to indicate the primary author, and the
% \aindex for all other co-authors.  You MUST use the following
% syntax: 
%
% SYNTAX:  \aindex{LASTNAME, F. M.}
% 
% where F is the first initial and M is the second initial (if
% used).  This guarantees that authors that appear in multiple papers
% will appear only once in the author index.  

\paindex{Schwarz, H. E.}
\aindex{Monteiro, H.}
%\aindex{Biemesderfer, C. D.}

%-----------------------------------------------------------------------
%                     Author list for page header
%-----------------------------------------------------------------------
% Please supply a list of author last names for the page header. in
% one of these formats:
%
% EXAMPLES:
% \authormark{LASTNAME}
% \authormark{LASTNAME1 \& LASTNAME2}
% \authormark{LASTNAME1, LASTNAME2, ... \& LASTNAMEn}
% \authormark{LASTNAME et al.}
%
% Use the "et al." form in the case of seven or more authors, or if
% the preferred form is too long to fit in the header.

\authormark{Schwarz \& Monteiro}

%-----------------------------------------------------------------------
%			Subject Index keywords
%-----------------------------------------------------------------------
% Enter up to 6 keywords describing your paper.  These will NOT be
% printed as part of your paper; however, they will be used to
% generate an object index and a subject index for the proceedings.  
% There is no standard list,  however, individual object names are
% encouraged and one or two word descriptions of the topics (e.g.MHD, 
% ionized gas) are useful. 
%
% EXAMPLE:  \keywords{NGC 7027, AFGL 2688, HD 161796, binary stars,
%                      dust,  molecular gas}
%


%-----------------------------------------------------------------------
%			       Abstract
%-----------------------------------------------------------------------
% Type abstract in the space below.  Consult the User Guide and Latex
% Information file for a list of supported macros (e.g. for typesetting 
% special symbols). Do not leave a blank line between \begin{abstract} 
% and the start of your text.

\begin{abstract}          % Leave intact
We show that there are strong links between certain types of
asymmetrical Planetary Nebulae (PNe) and symbiotic stars. Symbiotics
are binaries and several have extended optical nebulae all of which
are asymmetrical and $\ge$\,40\% are bipolar. Bipolar PNe are likely
to be formed by binaries and share many properties with symbiotic
nebulae (SyNe). Some PNe show point symmetry which is naturally
explained by precession in a binary system. M2-9 has both point and
plane symmetry, and has been shown to have a binary central object. We
show that inclination on the sky affects the observed properties of
bipolar nebulae due to enhanced equatorial densities, and compare
observations of a sample of BPNe with a simple model. Good agreement
is obtained between model predicted and observed IR/optical flux
ratios and apparent luminosities, which further confirms the binary
hypothesis.
\end{abstract}

%-----------------------------------------------------------------------
%			      Main Body
%-----------------------------------------------------------------------
% Place the text for the main body of the paper here.  You should use
% the \section command to label the various sections; use of
% \subsection is optional.  Significant words in section titles should
% be capitalized.  Sections and subsections will be numbered
% automatically. 

\section{Introduction}

Planetary Nebulae (PNe) are formed and live briefly at the spectacular
end of the lives of most low- and intermediate mass stars. The outer
layers of these stars are expelled when their cores collapse after
having used up their core fuel and switching off the nuclear furnace
that through radiation pressure kept gravity at bay. These tenuous
outer layers are then lit up by photo-ionization caused by the heating
up of the collapsing core. The White Dwarf (WD) that remains cools,
following a ``Sch\"{o}nberner track'' in the HR diagram on which the
cooling time depends critically on the WD mass, with massive stars
evolving much faster than lower mass objects. For PNe the average WD
mass is about 0.6\,M$_{\odot}$, with associated cooling time of about
10$^5$\,yrs. after which the star remnant fades away like a dying
ember.

This scenario was essentially predicted by Deutsch(1956) in a paper
called ``The Dead Stars of Population I''. He links the formation of
PNe to the last phase of red giants, and makes the connection with the
so-called ``combination variables'' which are now known to be the
symbiotic stars\footnote{The term''symbiotic
star'' was coined by Merrill\,(1950) and was not yet disseminated well
in 1956.}, systems with partially ionized gas and dust
surrounding a binary containing a hot, compact star and a cool giant.

The classical picture of a PN as a ``circle with dot in the middle''
has now been superseded by the rich complexity of observed
morphologies, and it is accepted that most PNe are asymmetrical. A
smaller fraction of PNe shows extreme asymmetries: the bipolar and
point-symmetric PNe. Related are the symbiotic nebulae (SyNe) which
have a high proportion of bipolar and other strongly asymmetric nebulae.

\section{Bipolars and Symbiotic Nebulae}

Bipolar PNe (BPNe) are a sub-set of the general group of PNe. There
are various definitions of what BPNe are but here I will use the one
formulated by Schwarz et al.\,(1993), noting that this is one of the
more restrictive definitions. A PN is bipolar when it has an aspect
ratio larger than unity and has a ``waist'' i.e. it has well defined
lobes and an overall dumb-bell shape. With this definition 12\,\% of the
objects in Manchado et al. (1996) are bipolar, 11\,\% in Schwarz et
al. (1992), and 9\,\% in G\'{o}rny et al.\,(1999), the three
major imaging catalogs of recent years. Therefore about 10\,\% of the
$\approx$\,600 PNe with good narrow-band images are truly bipolar.

Bipolarity is not just a morphological coincidence. Bipolar PNe have
properties that differ significantly from those of the general PNe
sample as was first reported by Corradi \& Schwarz\,(1995). They used
a sample of about 50 objects to show that: BPNe have a smaller scale
height (z\,=\,130\,pc vs. 260\,pc); hotter central stars
(mean\,=\,145\,kK vs. 75\,kK); nearer circular orbits in the Galaxy;
He, N, \& {\it Ne} overabundances; higher mean V$_{exp}$
(150\,kms$^{-1}$ vs. 15\,kms$^{-1}$; larger mean linear sizes
(0.76\,pc vs. 0.1\,pc); \& more massive progenitors ($\geq$
1.5\,M$_{\odot}$. vs.1.0\,M$_{\odot}$. Another piece of evidence comes
from Stanghellini et al.\,(1993) who investigated correlations between
the morphology of PNe and their CS loci in the HR diagram. They found
that bipolars have a different mass distribution of central stars from
those of elliptical PNe, and a lower ratio of HeII/HI Zanstra
temperatures than ellipticals (1.3 vs. 1.8) and therefore a different
optical depth. These results are based on a small sample of objects
but again physical properties correlate with morphology. We conclude
that BPNe come from a different population with different physical
properties.

Bipolars are among the most extreme asymmetrical PNe, something they
have in common with some of the SyNe that have been discovered. We
list the known SyNe in Table~1.

\begin{table}[!t]\centering
  \caption{Symbiotics with optical nebulae. Given are: name, type,
size of nebula, shape description, aspect ratio (AR), and the 
maximum expansion velocity. D and S are dusty and photospheric types 
resp. (Kenyon\,1986), Yell indicates yellow symbiotics with the G band 
in their spectrum.}
\vspace{0.3cm} 
\label{tab:symbs}
\begin{tabular}{lcrclr}
\hline Name & Type & Size (``) & Shape & AR & max\,2V$_{exp}$ \\
\hline 
AG Peg    & Yell & 8   & irregular      &     &     \\ 
AS 201    & Yell & 13  & elliptical     & 1.3 & 16  \\ 
BI Cru    & D    & 150 & bip.\,+\,jet   & 8   & 280 \\
CH Cyg    & S    & 32  & jet\,+\,irreg. & 1   & -   \\ 
H 1-36    & D    & 1.5 & unresolved     & -   & -   \\
H 2-2?    & S    & 1.4 & unresolved     & -   & -   \\ 
HBV 475   & S    & 0.4 & irregular      & -   & -   \\
He 2-104  & D    & 95  & 2 bip.\,+\,jet & 10  & 250 \\ 
He 2-147  & D    & 5   & proj. ring     & 2.2 & 100 \\ 
HM Sge    & D    & 30  & irregular      & 2.5 & -   \\ 
M2-9	  & D    & 115 & bipolar        & 12  & 328 \\
R Aqr     & D    & 190 & bip.\,+\,jet   & 6   & 200 \\ 
RX Pup    & D    & 4   & bipolar?       & -   & -   \\ 
V417 Cen  & Yell & 100 & bipolar        & 2   & 10  \\ 
V1016 Cyg & D    & 20  & elliptical     & 2   & -   \\
\hline
\end{tabular}
\end{table}

We put M2-9 in Table~1 since according to Balick\,1989 the object is
probably symbiotic. Note that 6 SyNe are bipolar or 40\% of the total,
cf. 10\% of PNe. He2-147 is possibly bipolar too, the outflow having
left the now visible ring. This would increase the bipolar SyNe
fraction to 47\%. This firmly links binarity with bipolarity. Note
that 5 out of the 7 SyNe with known expansion velocities have V$_{exp}
\ge$ 100\,km/s with an average V$_{exp}$\,=\,232\,km/s, similar to the
bipolars. Sa2-237 (Schwarz et al. 2002), M2-9 (Schwarz et al. 1997),
BI Cru (Schwarz \& Corradi 1992) and He2-104 (Corradi \& Schwarz 1995)
are all binaries, because these objects need a hot central source to
excite the observed [OIII] line but have low luminosities. Other
evidence shows that they have an evolved cool star (a Mira in He2-104
and BI Cru) so the central objects are symbiotic-like binaries. There
is also evidence that the bipolars A79, He2-428, and M1-91 are
binaries (Rodr\'iguez et al. 2001).

Summarizing: SyNe have 4 times more bipolars than normal PNe; average
V$_{exp}$\,=\,140\,km/s as for bipolars; average z\,=\,133\,pc as for
bipolars; bipolar SyNe have average z\,=\,98pc; most have [NII] as
strongest line as do bipolars. {\it So SyNe share many properties with
bipolar PNe and are known to be binaries.}

We note that the aspect ratio -which is a measure of the degree of
asymmetry or bipolarity- is correlated with V$_{exp}$ in
SyNe. Figure~1 plots this for the 7 SyNe for which both values are
known. One expects this behavior if the degree of collimation is
related to the outflow velocity.

\begin{figure}
\epsscale{.70}
\plotone{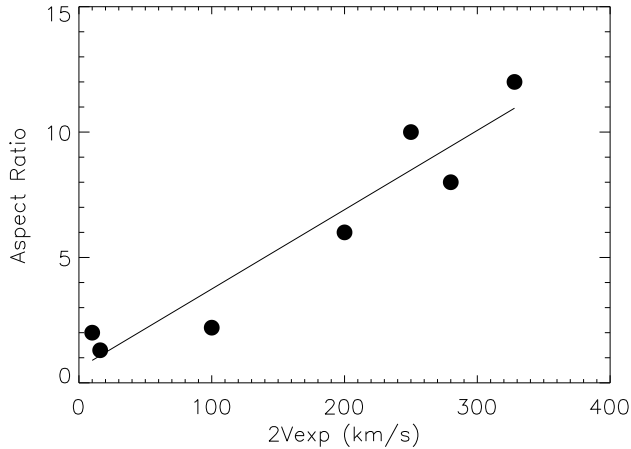}
\caption{The aspect ratio as a function of V$_{exp}$ for symbiotic nebulae. 
There is a strong positive correlation. The line is a linear regression to 
guide the eye.} \label{}
\end{figure}

Bipolar PNe have a preferred direction and this should produce
observable effects due to their orientation on the sky. In the next
section we investigate this.
 
\section{Inclination effects in bipolar nebulae}

PNe are randomly oriented in the Galaxy according to Corradi, Aznar \&
Mampaso\,(1998), and therefore the number distribution of observed
bipolar axes on the sky should follow a sin(i) law, where i is the
inclination to the line of sight. An inclination of 90$^{\circ}$ is
associated with a bipolar with its lobes in the plane of the sky,
0$^{\circ}$ means the lobes point toward the observer. If there is an
equatorial density enhancement in bipolars, as suggested by
e.g. Morris\,(1987), Icke, Balick, Frank\,(1992), Corradi
\& Schwarz\,(1995), Schwarz, Corradi, M\'endez\,(2002), then some effect
on their observed Spectral Energy Distributions (SEDs) and
luminosities is expected. The equatorial ``torus'' will produce an increased
extinction toward the central star when this is viewed at an
inclination nearer 90$^{\circ}$, because the stellar light passes
through more of the material. Some of the shorter wavelength light is
absorbed and re-emitted as FIR radiation, increasing the relative
contribution to the luminosity in the IRAS bands. Viewed pole-on, the
same object will show the central object, basically un-reddened plus
the torus, and the overall spectrum will be bluer.

The other mentioned effect is the variation of the luminosity with
orientation. Pole-on objects will be apparently over-luminous, due to
the fact that we see both the central star plus the re-radiated
emission from the torus, while equator-on nebulae will have a lower
observed luminosity since only the edge of the torus is seen. Random
inclination statistics assure that the mean luminosity over all
directions is constant and no energy conservation laws are violated.

We have collected a sample of 29 bipolars for which we have data
on the BVR, JHK, and IRAS fluxes, plus an inclination angle estimated
from optical images.  By plotting the relative luminosities in the BVR,
JHK, and IRAS bands (that is relative to the sum of the luminosities in those
three bands) we should see such effects, if they exist. The model
predicts that the IRAS luminosity should increase with the inclination
angle, and the other two bands should decrease. 

We made a simple model of a bipolar nebula: a star is surrounded by a
toroidal density distribution (``donut'') which absorbs \& re-radiates
15\% of the stellar flux. We then run this model for a random sample
of nebulae with their inclination angle histogram distributed on the
sky as sin(i). We generate binaries containing 100\,L$_{\odot}$ stars
with effective temperatures randomly distributed in the range
3800--6800\,K, and a 1000\,L$_{\odot}$ compact star with
25\,kK\,$\leq$\,T$_{eff}$\,$\leq$\,100\,kK with the equatorial torus
at 400\,K. The morphology and optical depth of the torus follow a
simple law. We generate fractional fluxes as for the observational
case and compare.

Figures\,2\,\&\,3 show respectively the observed distributions of
fractional fluxes and the model generated data. It is clear that the
model and observations are at least qualitatively in agreement. The
IRAS fluxes increase with inclination angle, the NIR, and visual bands
decrease. This lends strong support to the idea that bipolars indeed
have an equatorial density enhancement, and therefore are binaries.

\begin{figure}
\epsscale{0.5}
\plotone{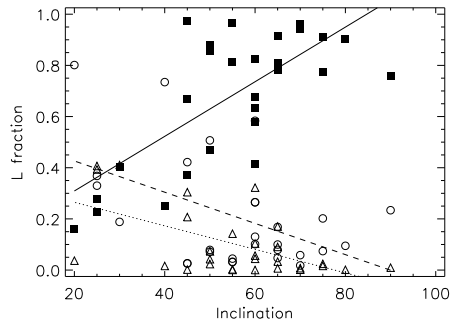}
\caption{The observed fractional 
fluxes in the visual (triangles; dotted line), NIR (circles, dashed line), 
and FIR (filled squares; solid line) as a function of inclination to the 
line of sight of a sample of bipolars. The lines are linear regressions.} 
\label{}
\end{figure}

\begin{figure}
\epsscale{0.5}
\plotone{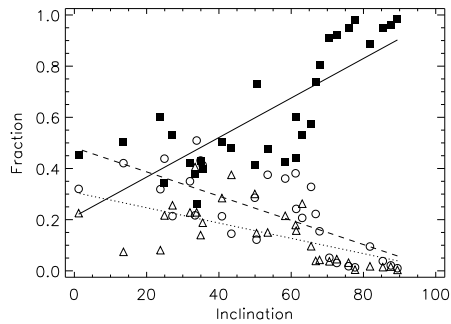}
\caption{The model generated fractional 
fluxes in the visual (triangles; dotted line), NIR (circles, dashed line), 
and FIR (filled squares; solid line) as a function of inclination to the 
line of sight of a sample of bipolars. The lines are linear regressions.} 
\label{}
\end{figure}

The luminosity of the model generated sample as a function of
inclination is shown in Figure~3, and shows the expected decrease with
inclination angle. Pole-on objects are super-luminous --since we see
the central object plus re-radiated emission from the torus-- while in
or near the plane of the sky they are sub-luminous (star absorbed and
only partially re-radiated toward the observer). 

\begin{figure}
\epsscale{0.6}
  \plotone{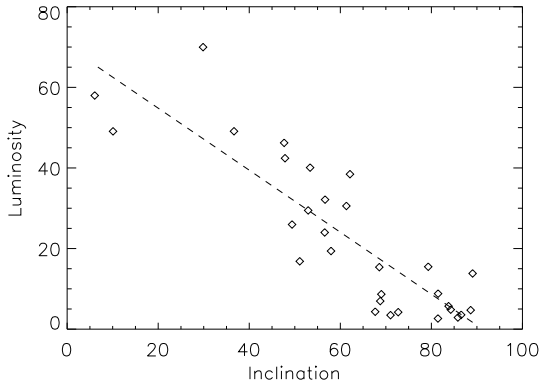}
  \caption{Luminosities (arb. units) of the model generated sample as a 
function of inclination.}
\end{figure}

An observational check of this model prediction is more difficult as
distances are not known to most objects. The few objects for which we
have reasonably hard distance determinations are listed in Table~1,
and they indeed show this effect. The small number of objects urge
caution when interpreting this.

There is a tendency for bipolars to be observationally selected nearer
the plane of the sky, i.e. with high inclination angles. This is due
to the fact that they have been mainly discovered by their morphology
and that is only recognized when they are not too far from the plane
of the sky, enhancing the sin(i) bias further. Bipolars should
therefore tend to have lower luminosities than the average for
PNe. This seems to be borne out by the observations, but also here,
care should be taken with small number statistics.

\begin{table}
  \caption{Observed luminosities and inclinations.}
  \begin{tabular}{lccc}
    \hline
    Name     & \multicolumn{3}{c}{Inclination $\geq$ 45$^{\circ}$} \\
    \hline
    Sa\,2-237 & i = 70 & 2.1 kpc &  340 L$_{\odot}$ \\
    M\,2-9    & i = 75 & 640  pc &  553 L$_{\odot}$ \\
    He\,2-104 & i = 50 & 800  pc &  205 L$_{\odot}$ \\
    He\,2-111 & i = 70 & 2.8 kpc &  440 L$_{\odot}$ \\
    M\,1-16   & i = 70 & 1.8 kpc &  194 L$_{\odot}$ \\
    \hline
    Name     & \multicolumn{3}{c}{Inclination $\leq$ 45$^{\circ}$} \\
    \hline
    R\,Aqr    & i = 20 & 200  pc & 2800 L$_{\odot}$ \\
    BI\,Cru   & i = 40 & 1.8 kpc & 3400 L$_{\odot}$ \\
    \hline
  \end{tabular}
\end{table}

In summary, bipolar PNe have much in common with SyNe; both types of
objects have binary central sources, and the concept of bipolars
having an equatorial density enhancement seems well established and
model data, based on randomly oriented nebulae, give a good fit to the
observations. The degree of asymmetry as measured by the aspect ratio
of the nebulae correlated positively with outflow velocity, as
expected if the collimation mechanism plays a role in defining the
maximum velocities.

\acknowledgments

HS thanks the organizers of APN3 for inviting him to give this
paper. HM acknowledges support from the NOAO science fund.

% Do not place any material after the references section

\end{document}